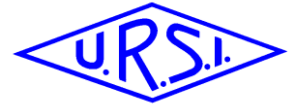

# An Evaluation of Distortion and Interference Sources originating Within a Millimeter-wave MIMO Testbed for 5G Communications

Tian Hong Loh* [1], David Humphreys[1], David Cheadle[1] and Koen Buisman[2]
(1) National Physical Laboratory, Teddington, Middlesex TW11 0LW, United Kingdom. http://www.npl.co.uk/
(2) Dept. of Microtechnology and Nanoscience/ Dept. of Electrical Engineering, Chalmers University of Technology, Gothenburg, Sweden

## Abstract

This paper presents an evaluation of distortion and interference sources, namely, the harmonic distortion and antenna crosstalk, originating within a 2 × 2 millimeter-wave (mm-wave) multiple-input-multiple-output (MIMO) testbed. The experience gained through the insight into the built testbed could be fed into the design of future mm-wave massive MIMO testbeds.

## 1. Introduction

In the fifth-generation (5G) wireless system, massive multiple-input-multiple-output (MIMO) and millimeter-wave mobile communications will have a significant role [1-4]. It is envisaged that massive MIMO base stations will utilise hundreds of antennas for communicating with multiple users. This could be achieved by spatial diversity where accurate channel state information (CSI) is required [1, 2]. However, the CSI quality and interference control will be critical factors to keep track of due to the simultaneous use of the same time-frequency resource for users also within the same cell and connected to the same base station.

The imperfect CSI and hardware imperfections will inevitably limit the system performance through inter-user interference. Furthermore, in-band interference sources such as intermodulation distortion (IMD) and mutual coupling between antennas or other parts of the analogue frontends will have a detrimental effect. The designer can spend fruitless efforts redesigning the system without insight of the actual cause, which can limit the uptake of these technologies. In this study, we evaluate interference and distortion sources originating within a 2 × 2 MIMO testbed operating at frequency in the mm-wave range [5]. The paper is organised as follow: Section 2 describes the system design, Sections 3 and 4 present the evaluation results for interference and distortion sources, respectively, and finally, conclusion are drawn in Section 5.

## 2. System Design

The testbed is a 2 × 2 mm-wave MIMO testbed and it is capable of performing spatial diversity MIMO transmission. The testbed hardware can be divided into two parts, namely, baseband, and radio frequency (RF) frontend. The baseband part was built using a pair of sub-6 GHz vector signal transceiver (VST) modules in a system with a real-time signal processing software defined radio (SDR) capability [6]. The RF frontend part was synthesized with several wideband off-the-shelf components and antennas (see the single-channel system layout in Figure 1) [5]. It consists of two pairs of standard gain horns [7] at the transmit- and receive-ends, the frequency up- and down-conversion hardware [8-12], and two independent local oscillators (LOs).

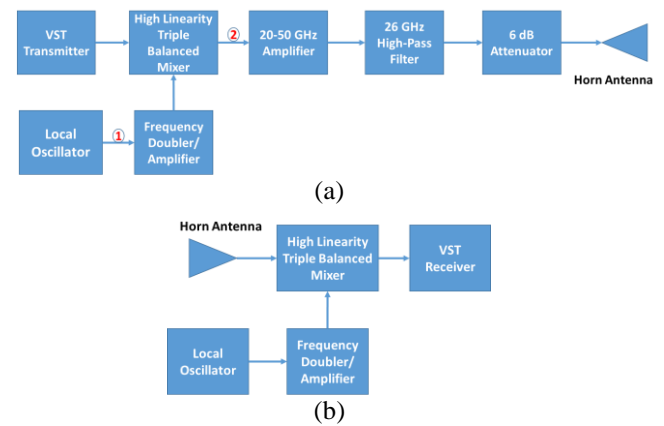

**Figure 1.** Single-channel layout of the MIMO system: (a) Transmit-end; (b) Receive-end.

At each single-channel transmit-end of the system, a microwave amplified frequency-doubling system [8] and a high-linearity triple-balanced mixer [9] were employed. Providing a suitable filter is chosen to limit spurious output, the system is envisaged to have operating RF frequency range from 20 up to 46 GHz with a sub-6 GHz baseband signal. At the receiver-end of the system the same components were used, except for the amplifier and filter, and is a mirror of the transmit-end configuration.

## 3. Distortion and unwanted RF sources

In this section, evaluation of several distortion sources of the system are presented. To operate the system at RF of

30 GHz, the LO and IF frequencies were chosen to be 25 GHz and 5 GHz, respectively [5].

### 3.1 Levelling response of the Amplified Frequency Doubler

The mixer [9] suggests a LO drive between +13 and +25 dBm and the doubler [8] input level specified as +5 to +10 dBm to achieve a doubled signal level of +20 dBm from 20 to 40 GHz. The behavior of the doubler over a range of input powers was not known. To investigate the above, the lower sideband (LSB) power level at the output of the mixer (marked as a circle with "2" shown in Figure 1(a)) was measured using a spectrum analyzer for a two-tone (10 MHz separation) IF signal at 5 GHz and a frequency input to the doubler of 12.5 GHz (25 GHz LO). Figure 2 shows the LSB power level variation plotted against the input power to the doubler (marked as a circle with "1" shown in Figure 1(a)). The results show that the amplified frequency doubler is levelled for LO power levels above – 12 dBm. Under normal operation we use a nominal LO power level of – 5 dBm to allow leeway for impedance mismatch and to remove the need for a high power-output synthesizer.

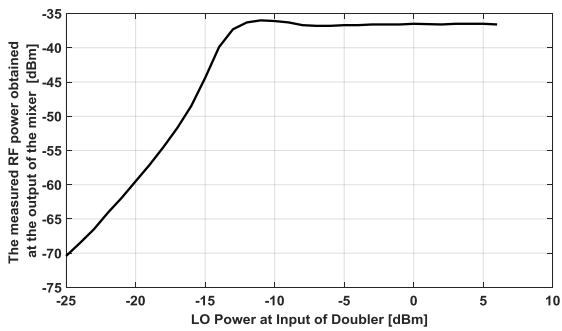

**Figure 2.** Levelling response of the amplified frequency-doubler.

### 3.2 LO Harmonic Distortion

In the design shown in Figure 1 [5], the same LO frequency was used for up-conversion and down-conversion. Inspection of the specification shows that in addition to the up-conversion to $f_{IF} + f_{LO}$, the mixer will produce signal components at higher frequencies $f_{IF} + 2f_{LO}$ and $f_{IF} + 3f_{LO}$. As the same LO and IF frequencies are used in both arms of the system, these components will appear as a degenerate sum at the IF frequency (see Figure 3(a)), adding to the signal error vector magnitude (EVM) [12]. By using different LO frequencies (see Figure 3(b)) these signal components can be separately identified (i.e. non-degenerate case).

Figures 4 and 5, show respectively, the system layout and the measured LO harmonic results for the non-degenerate case. A two-tone stimulus separated by 20 MHz at an IF frequency of 5.005 GHz and LO frequencies of 10.4975 GHz and 10.6975 GHz were used to give an RF frequency centered at around 26 GHz. Note that the frequencies were chosen to avoid the digital real-time oscilloscope (DRTO) sub-Nyquist spur frequencies. The measured results show that the $2f_{LO}$ and $3f_{LO}$ components are present at about 40 dB below the desired signals and identified in Figure 5. These components would therefore add an EVM contribution of about 1% to the result in the degenerate case [12] but can be removed using an appropriate band-pass filter.

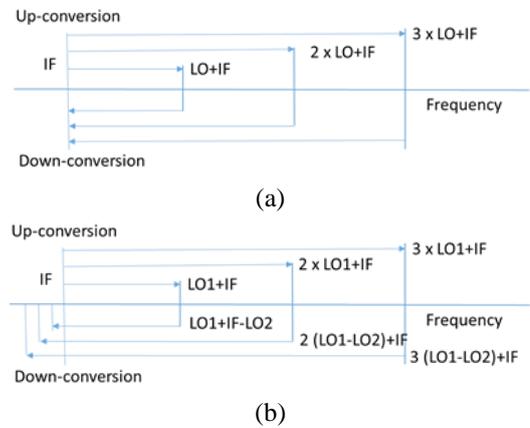

**Figure 3.** Frequency map of harmonically mixed components: (a) degenerate case; (b) non-degenerate case.

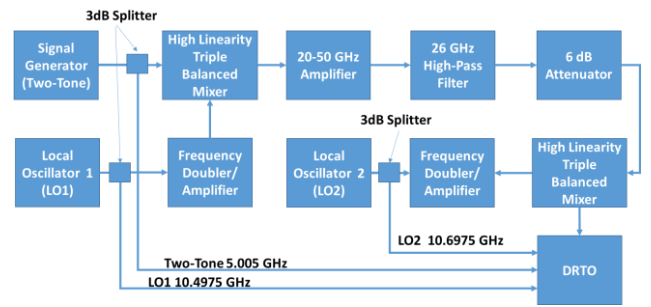

**Figure 4.** System layout for LO harmonic test.

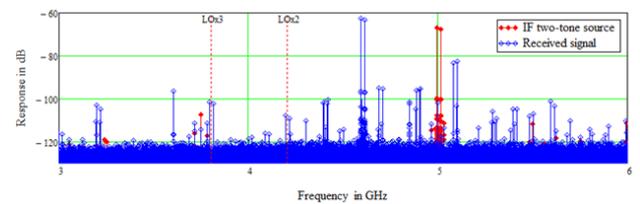

**Figure 5.** Results from non-degenerate test showing source and received signals.

### 3.3 Filtering unwanted signals

The initial link assessment results presented in [5], show that the roll-off of the 26 GHz high-pass filter provides insufficient isolation to suppress unwanted harmonic components. It is also sufficiently broad that the $2f_{LO}$ signals will be unfiltered. Any residual LO power at 25 GHz coupled through the mixer will also pass through the filter. In order to quantify the LO Breakthrough, the LSB, LO Breakthrough and upper sideband (USB) signal levels at the output of the filter were measured with a spectrum analyzer for an IF frequency centered at 5.005 GHz. The LO frequency was measured over 16 GHz to 31 GHz.

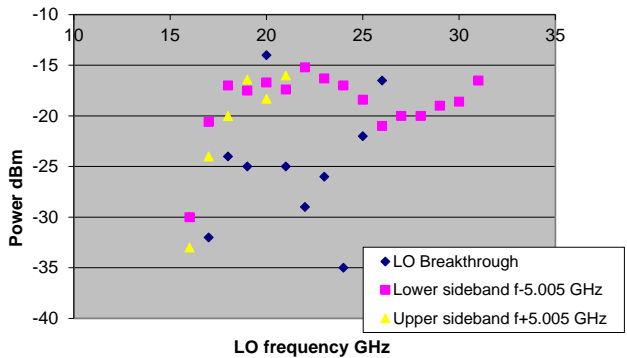

**Figure 6.** RF spectrum analyser display showing dominant LO breakthrough.

As depicted in Figure 6, the results show that the LO breakthrough varies significantly with frequency. Also, the power of the LO breakthrough is comparable to the desired signal for a LO frequency of 25 GHz and will limit the amplifier performance. To remove the unwanted harmonic and LO signals the broadband filter has been replaced with a commercial bandpass filter covering he range 27.5 GHz to 31 GHz [13]. The 6 dB attenuator has been repositioned between the mixer and the filter so that the level of the rejected LO and LSB signals is reduced.

### 3.4 Intermodulation

The residual components and nonlinearities have been measured for the system including the new filter using an RF power sweep of the 5.005 GHz two-tone signal (10 MHz separation) using the configuration shown in Figure 4. Figure 7 shows the levels of the first and second intermodulation terms (in-band) spaced at 10 MHz and 20 MHz from the main RF tones, and the residual components. The results show that the improved filtering removes the residual LO breakthrough term and the third order intercept power is 10.5 dBm and this nonlinearity is mainly attributable to the amplifier.

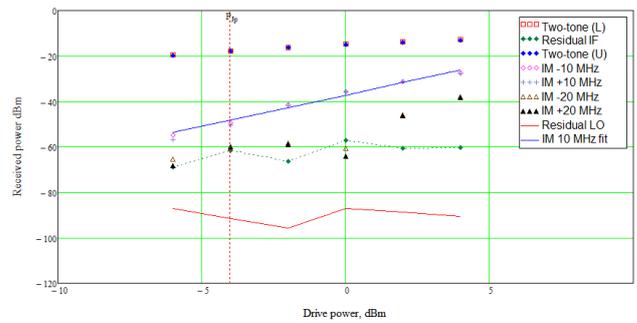

**Figure 7.** IF power sweep at 28 GHz RF frequency.

## 4. Evaluation of interference sources

In this section, an evaluation of interference sources due to antenna coupling is presented. To investigate the influence caused by antenna coupling in the system a pair of antennas were removed from the system and measured without the frequency conversion hardware attached. Measurements were conducted using a vector network analyser calibrated traceable to national standards. A 2.92 mm calibration kit was used to calibrate the system to the ends of the cables.

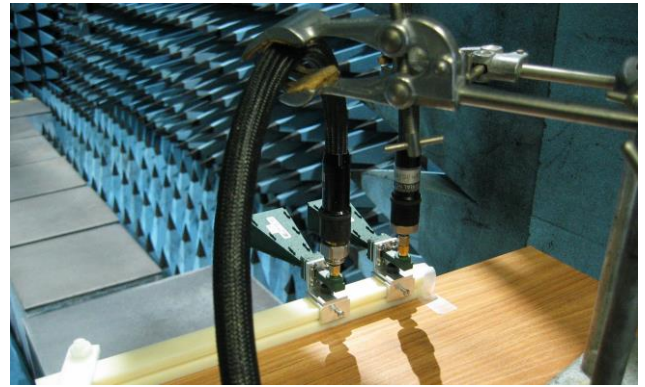

**Figure 8.** Experimental setup for the antenna coupling measurements in anechoic chamber at NPL.

The antennas were mounted in either a co-polarized or cross-polarized configuration using a system that allowed the separation between them to be adjusted (see Figure 8). Measurements were made with the separation between the outer edges of the horn antennas set between 1 mm and 41 mm in steps of 1 mm. The results shown in Figure 9 were measured at 30 GHz where the wavelength is approximately 10 mm. As shown in Figure 9, the coupling between of the directional standard gain horn antennas are insignificant in both configurations. Note however that the overall coupling for the co-polar configuration is slightly higher then cross-configuration. Also, the matching performance of the antennas has been observed. It is envisaged that this would introduce a difference in the link performance between the two channels in the MIMO system.

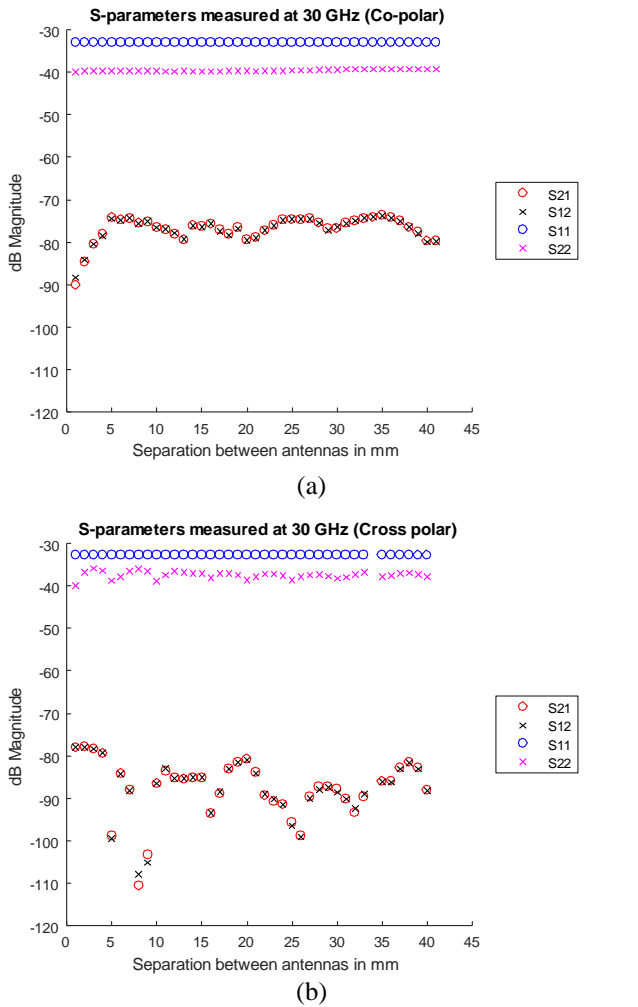

**Figure 9.** Measured *S*-parameters at 30 GHz for different antenna separations: (a) Co-polar configuration; (b) Cross-polar configuration.

## 5. Conclusions

This paper has presented an evaluation of interference and distortion sources originating within a 2 × 2 mm-wave MIMO testbed system, which offers a degree of flexibility that enables the investigation on the signal test, communication algorithm and measurement metrology for 5G communications. This work enables determination of possible points of weakness for potential future 5G system.

## 6. Acknowledgements

The work was supported in part by the 2017 – 2020 National Measurement System Programme of the UK government's Department for Business, Energy and Industrial Strategy (BEIS), under Science Theme Reference EMT17 of that Programme and in part by the EU project MET5G – Metrology for 5G communications (this project has received funding from the EMPIR programme co-financed by the Participating States and from the European Union's Horizon 2020 research and innovation programme), under EURAMET Reference 14IND10.